\documentclass[aps,prb,twocolumn,showpacs,letterpaper]{revtex4-1}
\usepackage{graphicx}
\usepackage{subfigure}
\usepackage{amsmath}
\usepackage{color}

\newcommand{\cfa}{{CaFe$_{2}$As$_{2}$}}

\newcommand{\mfa}{{MFe$_{2}$As$_{2}$}}
\newcommand{\mcfaq}{{Ca$_{0.75}$M$_{0.25}$Fe$_2$As$_2$}}

\begin{document}

%================================
% TITLE
%================================
\title{First-principles study of superconducting Rare-earth doped CaFe$_2$As$_2$ } 

%================================
% LIST OF AUTHORS  
%================================

\author{A.~Sanna$^{1,2}$, G.~Profeta$^1$, S.~Massidda$^3$ and  E.K.U.~Gross$^{1,2}$}

\affiliation{$^1$ Max-Planck Institute of Microstructure Physics, Weinberg 2, D-06120 Halle, Germany}
\affiliation{$^2$ European Theoretical Spectroscopy Facility (ETSF)}
\affiliation{$^3$ Dipartimento di Fisica, Universit\`a degli Studi di Cagliari, I-09042 Monserrato, Italy}

\begin{abstract}
We report a systematic and ab-initio electronic structure calculation of \mcfaq\ with M = Ca, Sr, Eu, La, Ce, Pr, Nd, Pm, Sm, Na, K, Rb. 
The recently reported experimentally observed structural trends in rare earths-doped \cfa\ compounds are successfully predicted  and a complete theoretical description of the pressure induced orthorhombic to collapsed tetragonal transition is given. We demonstrate that the transition pressure is reduced by electron doping and rises linearly with the ionic size of the dopants. We discuss the implications of our description for the realization of a superconducting phase.

\end{abstract}
\pacs{   }

\maketitle

The family of iron based superconductors (SC) continues to grow with the discovery of new systems that add to the hundreds already known\cite{reviewpaglione,reviewmazin, reviewstewart}. However, in spite of this enormous amount of experimental data, an ab-initio theory describing these superconductors is still missing, and the search for new materials is guided only by some observed features that appear to be in common.

Even if it is not a general rule, the so-called parent compounds do not superconduct without doping\cite{reviewstewart}
 or without applying  high pressure\cite{reviewChu2}.
The rationale for this behavior could be naively ascribed to the necessity to destabilize the orthorhombic (O) antiferromagnetic order of the parent compounds to promote a magnetic mediated pairing field\cite{Mazin}.
Indeed, superconductivity is realized in tetragonal (T) phases only after the magnetic order is suppressed, although some coexisting phases were discovered in 122 compounds\cite{Chu}.
 
In the so-called 122 family (\mfa, M=Ca, Ba, Sr)  superconductivity can be induced under high pressure\cite{reviewChu2}. The critical pressures at which superconductivity is detected varies with M\cite{Torikachvili,Alireza} and with the pressure conditions\cite{Mittal, Gianni1, Gianni2, Colombier, Kimber, Uhoya}  (hydrostatic and non-hydrostatic).
At the same time, the members of the 122 family show a pressure induced structural phase transition to a \textit{collapsed tetragonal} (CT) phase\cite{Goldman,Pratt}.
The main experimental evidence of the CT phase transition is the sudden decrease of the $c$ lattice parameter and a subsequent increase of the in-plane $a$ lattice constant.
First-principles calculations were able to describe this phase transition as induced by the formation of a direct As-As bond along the $c$-axis of the tetragonal phase \cite{Gianni2,Yildirim}. The CT phase is predicted to be a non-magnetic phase (the magnetic moment being zero on the Fe sites) in agreement with experimental results\cite{Goldman} which report the disappearance of the magnetic response.

In Ba and Sr 122 compounds, where the O-T and T-CT are well separated in pressure\cite{Alireza,Mittal}, it seems that the superconducting phase is realized across the O-T transition (even if the role of non-hydrostaticity of the pressure medium is not clear), while it was debated in \cfa\ where the transitions are very near in pressure\cite{Kreyssig} and hysteresis effects are relevant.
In fact, for \cfa\ the  precise nature of the crystallographic structure of the superconducting phase is under debate, but the CT phase obtained under hydrostatic pressure conditions\cite{Kreyssig,Prokes,Yu} seems to be ruled out. 
This suggests that superconductivity is realized in a tetragonal paramagnetic structure (not collapsed) realized only under non-hydrostatic conditions\cite{Pratt}. This phase should be  characterized by the proximity of a magnetic instability absent in the CT phase.

This subject was revived  with the discovery of high temperature superconductivity (up to 45-49 K) in \cfa\ doped with trivalent rare-earth (RE) metal atoms (La, Pr, Ce, Nd) substituting the divalent Ca atoms\cite{Qi,Lv,Goh,Saha,Gao}. 
%Lower $T_c$ were reported as effect of non-RE-doping\cite{Wu}.
It was demonstrated that rare earths have high solubility (up to 27\% in Ref.~\onlinecite{Saha}) in \cfa\ mainly due to close matching between ionic-radii of RE and Ca, and that their incorporation suppresses the low temperature AFM ordering in favor of the superconducting phase.
Application of hydrostatic pressure on the doped materials is shown to increase further the critical temperature\cite{Goh}.

Moreover, neutron diffraction experiments as a function of temperature revealed an interesting structural property: depending on the RE dopants, the low temperature phase shows the characteristics of the CT phase\cite{Saha}.
In fact,  some of the systems (Pr, Nd doped for example) exhibit a low temperature phase in which the out-of-plane axis shrinks of about 10\% its high temperature value. 
On the other hand, when doped with La or Ce, this last structural phase transition is not observed. Nevertheless, superconductivity is found irrespective of the presence or not of the O-CT transition.

Recently the scenario was further enriched with the reports of superconductivity in Ca$_{0.67}$Sr$_{0.33}$Fe$_2$As$_2$ at high pressure\cite{pagl_sr} and of quenched Fe magnetic moment in the collapsed tetragonal phase of Ca$_{1-x}$Pr$_{x}$Fe$_2$As$_2$\cite{pagl_pr}.

At the moment many questions remain to be answered, in particular regarding the origin of the superconducting phase.
In addition, the details of the structural and electronic properties of these new phases are still unknown and it is not clear if the magnetic correlations are suppressed/enhanced upon rare-earth doping in the T phase and the cT phase\cite{pagl_pr}.
As a first step in the comprehension, we report  first-principles DFT\cite{methods} prediction of  structural, magnetic and electronic properties of \cfa\ upon doping with RE. In addition, we extended the study to isovalent alkaline earths and Eu$^{2+}$ and consider the effect of hydrostatic pressure on doped systems.

We start our discussion considering the physics of the parent compound, \cfa. 
Density Functional Theory successfully predicts the structural collapse of the tetragonal phase at a pressure P$_{{\rm O}\to{\rm CT}}$ of 0.8 GPa, as calculated from the crossing point of the free energy of the two phases (the O and CT).
This value is in reasonable agreement with experimental evidences of O-CT phase transition at $P> 0.35$ GPa\cite{Kreyssig}. 
The difference is in part due to computational accuracy, to the GGA functional and experimental uncertainties\cite{Kreyssig, Goldman, Yu}.
The existence of both O and CT solution at the same pressure, means that the two phases are locally stable. This agrees with the first order nature of the phase transition and suggests possible experimental hysteresis effects, as indeed reported\cite{Goldman}.

\begin{figure}[h]
\includegraphics[width=0.45\textwidth]{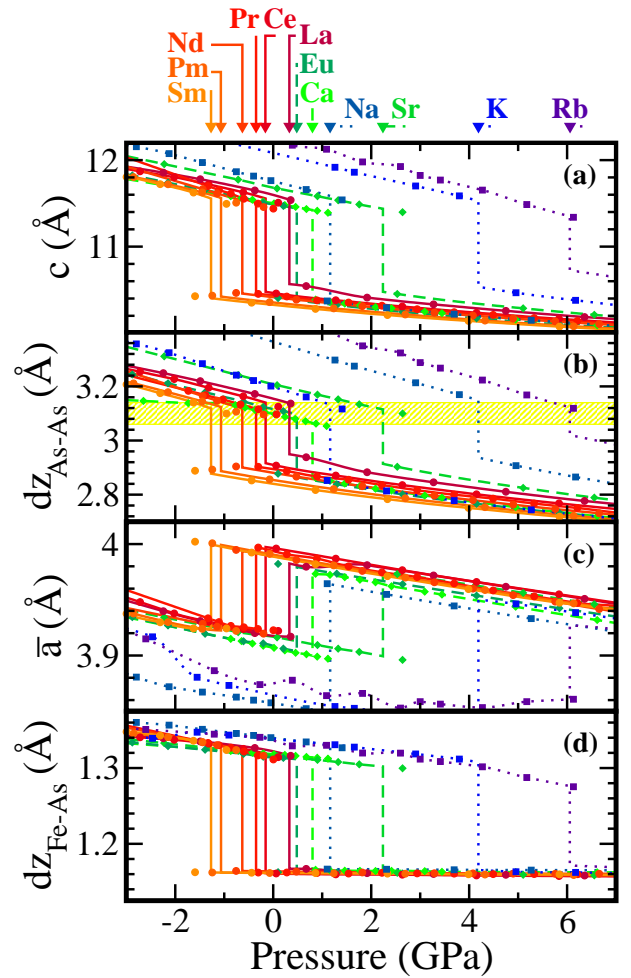}
\caption{(color online) Pressure dependence of the theoretical lattice parameters of \mcfaq\ in the CT and O phases. 
The curves follow the phase that minimizes the enthalpy as a function of P. The data points that lie off from the lines are locally stable solutions with higher enthalpy.
The structural transition is marked by a collapse of the $c$ lattice parameter (see panel $a$) and an expansion in the $ab$ directions ($\bar {\rm a}$ denotes the average between the two in plane axis parameters) as shown in panel $c$.  $dz_{Fe-As}$ is distance between As an Fe layers, and $dz_{As-As}$ is the interlayer distance between As atoms.
}\label{fig:structure}

\end{figure}

The calculated CT lattice constant (see Table~\ref{tab:data} and Fig.~\ref{fig:structure})  are $a=3.97 $ \AA, $c=10.38 $ \AA\ (near the transition pressure), to be compared with experimental values 3.99 \AA\ and 10.62  \AA.  
The O phase has $a=3.86$ \AA,  $ b=3.94 $ \AA\ and $c=11.41 $ \AA, while the experimental value\cite{Kreyssig} is $a=3.87 $ \AA, $ b= 3.91 $ \AA\ and $c= 11.65 $  \AA. 

The structural collapse reduces the $c$ lattice parameter by about 8\%, while the $a$ and $b$ lattice parameters have a expansion of $\sim 1.8$\% (as an average between the two orthorhombic directions).
The compressibility also changes abruptly. We estimate a bulk modulus ($-V\frac{\partial P}{\partial V}$) jump across the transition of $\sim 20$ GPa. 
The structural transition into the collapsed phase is also accompanied with a magnetic to non-magnetic transition in which the magnetic moment at the iron site drops from $\sim 1.35 \mu_B$ to zero. Contrary to the low pressure tetragonal phase, 
the CT phase is far from any magnetic instability (we could  not converge to any magnetic solution), in agreement with the experimental observation of no spin-fluctuations in th CT-phase\cite{Pratt}.
This is indeed the observed phenomenology of the O-CT phase transition, that is correctly captured by the DFT-GGA. 
 
We switch now to the description of doped \cfa, focusing our interest on those systems that experimentally show enhanced superconducting properties. 

One of the main experimental evidences reported in Ref.~\onlinecite{Saha} is the observation of 
 a clear discontinuity on the variation of the lattice constants ($a$ and $c$) as a function of the temperature for  15\% doped Praseodymium, 
 pure \cfa\ and 8\% doped Neodymium (but not in Lanthanum doped). 
 
In order to predict the structural changes upon chemical doping and to understand the experimental results, 
we simulated the partial chemical substitution at the Ca site with three different dopants:  isovalent (Sr and Eu), aliovalent with trivalent electron-dopants  RE (the experimentally realized case La, Ce, Pr, Nd and the not yet realized Pm and Sm) and hole-doping with alkali Na, K, Rb. 
The doping level is fixed to x=25\%, well within the experimentally realized doping range in \cfa\ \cite{Saha, pagl_sr}. 
We performed structural optimization at different volumes in both non-magnetic and antiferromagnetic (stripe) phase. 
Phase stability was calculated comparing enthalpy curves as a function of the external pressure.
Structural, electronic and magnetic properties are collected in Tab.~\ref{tab:data}. 

\begin{table}[h]
\begin{tabular}{l|ccccc}
 M$\;\;\;$        & $\;\;P_{O\to CT}$       &  $\;\;\;\;\;V_0\;\;\;\;\;$    &  $\;\;\;\;V_{CT}\;\;\;\;$   &$\;\;\;\;\Delta E\;\;\;$ &$M_{Fe}\;$ \\ \hline
  Ca       &         0.80                 &        87.79            &        82.78         &         3.63      &         1.45     \\
  Sr       &         2.24                 &        89.60            &        84.86         &         9.27      &         1.34     \\
  Eu       &         0.48                 &        87.75            &        82.59         &         2.26      &         1.37     \\
  La       &         0.33                 &        88.99            &        84.08         &         1.47      &         1.43     \\
  Ce       &        -0.16                 &        88.64            &        83.55         &        -0.74      &         1.40     \\
  Pr       &        -0.35                 &        88.10            &        83.23         &        -1.50      &         1.42     \\
  Nd       &        -0.63                 &        87.71            &        82.87         &        -2.95      &         1.44     \\
  Pm       &        -1.07               &        87.40          &        82.44           &        -5.22      &         1.45     \\
  Sm       &        -1.27                &        87.07         &        82.15           &        -6.29      &         1.45     \\
  Na       &         1.16                 &        87.43        &        82.54           &         4.99      &         1.22     \\
  K        &         4.19                  &        90.19        &        86.57           &        12.66      &         1.14     \\
  Rb       &         6.06                &        92.08        &        88.61            &        14.78      &         1.10     \\

\end{tabular}
\caption{Structural electronic and magnetic properties of M doped \cfa. 
$P_{O \to CT}$ (GPa) is the transition pressure of the structural collapse.
$V_O$ and  $V_{CT}$ (\AA$^3$) are the equilibrium volumes in each phase, 
$\Delta E$ (meV/f.u.) is the energy difference between the two phases at zero pressure.  
$M_{Fe}$ ($\mu_B$) represents the magnetic moment in the O phase.
}
\label{tab:data} 
\end{table}

The characteristic transition pressure P$_{{\rm O}\to{\rm CT}}$ from the orthorhombic to the collapsed tetragonal phase as obtained in the pure \cfa\ is strongly affected by the chemical substitution. 
In particular, Ce, Pr, Nd, Pm and Sm are stable in the collapsed phase already at ambient (zero)  pressure, 
while a positive pressure is needed to induce a collapse in Sr, Eu, La, Na, K and Rb  substituted systems. 
This behavior is in excellent agreement with the available experimental data reported by Saha and coworkers \cite{Saha, pagl_sr}. 

In fact, Pr is reported to induce the structural collapse already at a concentration between 5\% and 7.5\% ; Nd is collapsed at 8\% of doping, while no collapse is reported by La doping up to 27\% of doping. 
In particular, we nicely predict the existence of the CT phase at 2.24 GPa in Sr doped \cfa\, as experimentally observed\cite{pagl_sr}.

On the contrary, Ce is not observed in the collapsed tetragonal phase up to a 22\% of content, contrary with the theoretical predictions. However, Ce was predicted to be at the boundary of the phase transition, in fact we find the  collapsed phase only very slightly more stable than the O phase by 0.7 meV/f.u. and is sufficient a tiny positive pressure to induce it (see Fig.~\ref{fig:structure}).

The evolution of the crystal structure upon doping and pressure is reported in Fig. \ref{fig:structure}. The data from the simulations on the two phases have been connected with a spline fit jumping at the calculated transition pressure. 
Lattice constants, as a function of pressure, show the characteristic drop of the $c$-axis (see Fig.~\ref{fig:structure}(a) ). 

An excellent agreement between theoretical and experimental lattice parameters (not shown) is observed when comparing the low-temperature experimental lattice constants\cite{Saha} with the calculated ones in the non-magnetic tetragonal phase (CT), confirming the validity of DFT in the description of this phase.
On the other hand, the high temperature phase is a magnetic (probably paramagnetic tetragonal) phase, whose description is improved, within DFT-GGA, including magnetic polarization of the Fe sites. 

The internal relaxation shows an interesting pressure dependence of the Fe-As interlayer distance ($dz_{Fe-As}$). In the O phase it decreases monotonically with the applied pressure while it becomes constant and material independent in the CT phase (as shown in Fig.~\ref{fig:structure}(d) ). Due to this feature in this second phase the lattice parameter $c$ depends only on the size of the impurity atom and can be considered a measure of its ionic radius in this class of compounds.

Being originated by the formation of As-As bond, the phase transition to the collapsed phase happens when the interlayer As-As separation ($dz_{As-As}$) reaches the critical value of 3.0\AA\ (about twice the As covalent radius)\cite{Gianni2}, independently of the dopant (shown in Fig.~\ref{fig:structure}(b)).
Indeed, this last result is confirmed by  X-rays diffraction measurements of Saha and coworkers which suggested\cite{Saha} that the key parameter that controls the collapse is the As-As distance. 

\begin{figure}[h]

\includegraphics[clip=,width=0.5\textwidth]{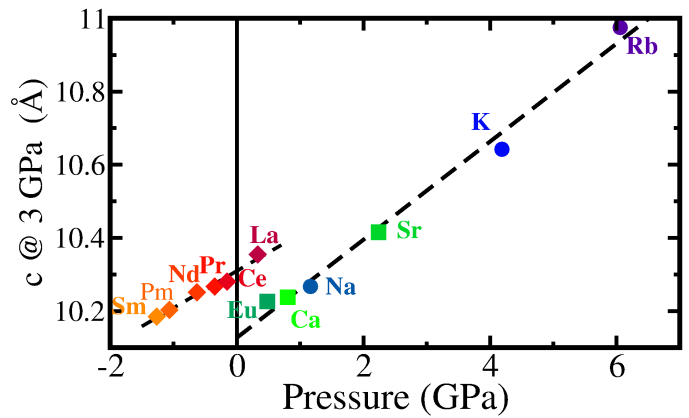}
\caption{(color online)  Lattice parameter $c$ versus $O\to CT$ transition pressure for each dopant in \mcfaq (M is indicated as a label for each point).  Dashed lines are guide to the eyes.
}\label{fig:c_P}
\end{figure}

Then, if we look at the correlation between the $c$ lattice parameter (in the CT phase) and the transition pressure P$_{{\rm O}\to{\rm CT}}$, we find a very interesting correlation. For all trivalent lanthanides the transition pressure scales linearly with the value of the $c$ lattice ($i.e.$ with the ionic size). The same happens for bivalent and mono-valent dopants, although with different linear coefficients, see Fig.~\ref{fig:c_P}.

This clearly tells that  P$_{{\rm O}\to{\rm CT}}$ is determined by  the size of the impurity atom and 
 the kind and level of doping, with electron doping favoring the CT transition; 

In fact, a smaller ionic size favors the structural collapse because when As atoms get closer their interaction becomes stronger and the required pressure to induce the O-CT transition is low (or zero).

\begin{figure}[htp]
\begin{center}
\includegraphics[clip=,width=0.45\textwidth]{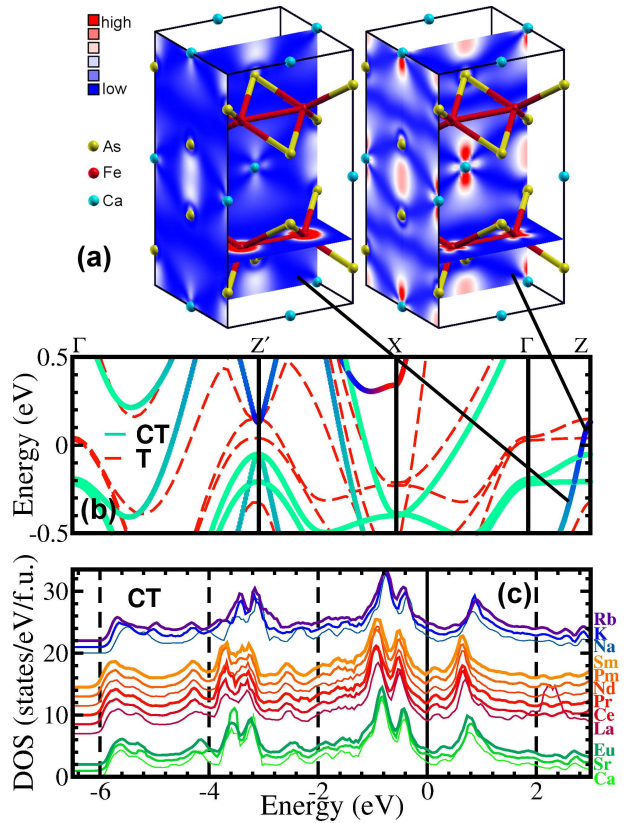}
\end{center}
\caption{(color online) Band structure of collapsed (CT) and uncollapsed (T) \cfa\ (panel b). The color scale used in the CT bands indicates the projection on As-$p_z$ states (blue for high projection, green for low). On the top (panel a) is shown the periodic part of two selected electronic wavefunction (indicated with black lines). 
Doping dependence of the DOS in the CT phase (panel c). An arbitrary vertical shift has been applied to improve the readability of the figure.
} 
\label{fig:bands}
\end{figure}

The valence of the dopant, on the other hand, also affects the transition pressure. 
We observe that the hole doping induced by the alkali is mainly located inside the Fe-As layer, not affecting sensibly the As-As vertical bonding energy, while electron doping fills the As-As bonding states,  favoring the chemical bonding. 
To support this picture we plot in Fig.~\ref{fig:bands}(a) the periodic part of two states belonging to the same electronic band but taken above and below the Fermi energy (indicated in figure). The interlayer bonding character is stronger above than below,  therefore while electron doping contributes to the interlayer bonding, hole doping affects it only to a little extent.

At this point we switch to the second part of our analysis. We focus on the electronic properties and their doping dependence. 
The non-magnetic band structure of undoped \cfa\ in the bct unit cell is plotted in Fig.~\ref{fig:bands}(b). 
We observe that both O (not shown) and CT phases show a depletion of states at the Fermi energy, with respect to this non-magnetic high-temperature phase. In the former case this is due to a magnetic splitting, while in the latter to chemical splitting of bonding and antibonding $p_z$ state in As. 

Considering the O phase, the electronic effects of aliovalent atoms can not be predicted by means of a  rigid band approximation because of the doping dependence of the magnetic moment (see Tab.~\ref{tab:data}), that strongly affects the band structure near the Fermi energy. 
On the other hand in the CT phases of all the investigated systems (near the critical lattice constants) a rigid shift of the Fermi level  reproduces very well the dopant induced modifications of the electronic structure. This is clearly seen comparing the DOS of different compounds as shown in the lower panel of Fig.~\ref{fig:bands}. 

Thus, apart from a small ($\simeq 0.1$ eV) rigid shift of the Fermi level, 
the non-magnetic band structure of the compressed tetragonal phase of \cfa\ (shown in 
Fig.~\ref{fig:bands}) is a good approximation for that of (superconducting) RE-doped \cfa\ \cite{comment}. 

The structural collapse induced by the dopants  profoundly changes the topology of the Fermi surface with respect to the non-magnetic tetragonal phase as it does in undoped \cfa.
The two-dimensional, cylindrical, hole FS's at the $\Gamma$ point disappear upon compression and three dimensional FS appear as a consequence of band crossing the Fermi energy along the $\Gamma$-$Z$ direction of the Brillouin zone. Thus,
in the CT phase it is very difficult to recognize the characteristic topology of FS common to other pnictides superconductors, and that are thought to be the basic ingredient of the SC pairing (namely the nearly nested cylindrical Fermi surfaces).
In addition, the magnetic moment of Fe site is predicted to be completely quenched  in the CT as recently confirmed by NMR $^{75}$ As spectra in Pr doped \cfa\ \cite{pagl_pr}. 

A possible scenario to reconcile the theoretical description and the experimental evidences  could be the same proposed by Prok\v{e}s and coworkers\cite{Prokes} to explain the superconducting phase in pure \cfa\ under pressure: superconductivity can  exist because of the formation of a new tetragonal phase (stabilized by chemical and/or strain/non-hydrostatic effects) which prevents both the magnetic and the structural phase transition, then retaining the characteristic electronic properties that makes it unstable toward superconductivity.
We, thus, call for further experiments aimed to characterize  the electronic, magnetic and structural properties of these new superconductors, in order to elucidate the nature of the non-collapsed tetragonal phase

Within the parameter-free DFT-GGA computational framework we explain the observed low-temperature and pressure 
phase transitions  observed experimentally in RE-doped \cfa\ and we predict, giving a rationale, the pressure dependent structural trends.

We show that, when realized, the low temperature CT phases induced by RE substitutions are indeed non-magnetic, with electronic band structures sharing the same features of the band structure of the CT phase of pure \cfa. Characteristic features promoting the superconducting phase are missing.
For this reason, the superconducting instability, if it is confirmed to be present in the compressed phases of RE-doped \cfa\, seems to open a new chapter in the physics of iron-based superconductors. However a scenario that we find more likely is that superconductivity appears in a non collapsed subphases as it was suggested by Pro\v{k}es and coworkers for undoped \cfa.

\section{Acknowledgments}
S.M. acknowledges support by the Italian MIUR through Grant No. PRIN2008XWLWF9.


\begin{thebibliography}{99}
\bibitem{comment}
With this model system, we are indeed neglecting possible disorder effects and the role of the electron dopant.
However, the model should grasp the main electronic properties of the compressed system.
\end{thebibliography}
\end{document}